\documentclass[twocolumn,showpacs,preprintnumbers,amsmath,amssymb]{revtex4}
\usepackage{graphicx}
\usepackage{dcolumn}
\usepackage{bm,color}

\def\be{\begin{equation}}
\def\ee{\end{equation}}
\def\bea{\begin{eqnarray}}
\def\eea{\end{eqnarray}}

\def\md{\mathrm{d}}

\def\s{\sigma}

\begin{document}


\title{The complete study on the inclusive production of $\Upsilon+\gamma$ at the LHC}



\author{Rong Li$^{1,3}$ and Jian-Xiong Wang$^{2,3}$}
\address{$^1$School of Science, Xi'an Jiaotong University, Xi'an 710049,
China.\\
$^2$Institute of High Energy Physics, Chinese Academy of Sciences,
P.O. Box 918(4), Beijing, 100049, China.\\
$^3$Theoretical Physics Center for Science Facilities, CAS, Beijing,
100049, China.}



\date{\today}

\begin{abstract}

In this paper we investigate the inclusive associated production of
$\Upsilon+\gamma$ at the hadron collider. We calculate the
color-singlet sub-processes and all three color-octet sub-processes
at the next-to-leading order. Seven sets of long distance matrix
elements(LDMEs), which are extracted from the studies on the prompt
production of $\Upsilon$ at hadron colliders, are used to give the
numerical results and we find that there are three sets of LDMEs
give the unphysical results on the yield and the polarization for
this process. The yield is enhanced by several or two orders times
comparing to that of color-singlet process and the polarization
changes from longitudinal to slightly transverse or even mainly
transverse in the large $p_t$ region. The estimation results
indicate that to study the process at the large hadron collider (LHC) 
is difficult and it may be well investigated at the future Super Proton-Proton
Collider(SPPC).
\end{abstract}


\maketitle


The production and decay of heavy quarkonium have become an active
area in physics since its discovery in 1974. Although there are some
defects in the theoretical description, the color singlet model(CSM)
had been the main tool to study the heavy quark system for about 20
years. In the early 1990s, in order to solve the $J/\psi$ surplus
problem at the Tevatron~\cite{Abe:1992ww} the non-relativistic
quantum chromodynamics(NRQCD) had been
proposed~\cite{Bodwin:1994jh}. This new scheme also factorizes the
physical process into the short distance part, which can be
calculated perturbatively, and the long distance matrix
elements(LDMEs), which can be extracted from matching the
theoretical prediction and the experimental data or calculated using
the lattice method. The extensive studies on the NRQCD and the heavy
quarkonium can be found in the review
references~\cite{Andronic:2015wma}.

In reference~\cite{Butenschoen:2012px} the authors had noticed that
the LDMEs of $J/\psi$ extracted from the different process can not
be consistent with each other. This may raise the doubt on the
universality of the LDMEs in the NRQD factorization scheme. As for
the $\Upsilon$, the heavy bottom quark mass makes it a better one to
be described by the NRQCD and the dilepton decay channel of it also
provides great convenience for experimental investigation. In 2007
in the color-singlet framework the yield of inclusive production of
$\Upsilon$ at next-to-leading order(NLO) was presented in
Ref.~\cite{Campbell:2007ws} and the polarization distribution was
obtained in the next year~\cite{Gong:2008hk}. The authors also
estimate the results on the yield and the polarization at the
partial next-to-next-to-leading order~\cite{Artoisenet:2008fc}.
Although these works are impressive progress which enhanced the
$p_t$ distribution largely in the large $p_t$ region and changed the
polarization from transverse to longitudinal one in the CSM the
theoretical predictions without the NLO results of color octet
channel hardly can be proper results to compare with the
experimental data. Neither the leading order(LO) results of CSM or
NRQCD nor the NLO results of CSM can reconcile the conflict between
theoretical results and the data on the $p_t$ distribution and the
polarization simultaneously. In order to investigate the heavy
quarkonium production at NLO in NRQCD we need not only to calculate
the short distance parts at NLO but also to determine the value of
corresponding LDMEs from matching the theoretical result with the
experimental data. The NLO results of S wave color octet channel are
calculated in Ref.~\cite{Gong:2010bk}. The first theoretical
calculation including the full NLO results of the color octet
channel is presented in Ref.~\cite{Wang:2012is}. There are some
other works on the full NLO theoretical investigation of inclusive
hadroproduction of
$\Upsilon$~\cite{Gong:2013qka,Han:2014kxa,Feng:2015wka}. By using
different schemes on the chosen of NRQCD scale or the fitting
procedure several sets of LDMEs were obtained. Although the relative
satisfactory results on the yield and the polarization prediction
were given in the above references there are large numerical
difference among different sets of LDMEs.

Recent years, except the inclusive production of heavy quarkonium
the associated production of quarkonium had also been an attractive
topic. From 1992 some authors had studied the haroproduction of
quarkonium associated with a gauge boson, photon, $W^{\pm}$ or $Z^0$
to investigate gluon content of the proton~\cite{Drees:1991ig} or
the color octet mechanism in the NRQCD at tree
level~\cite{Kim:1994bm}. The QCD NLO correction to some of these
process had been studied in Ref.\cite{Li:2010hc}. The color singlet
channel of $pp \to \Upsilon+\gamma+X$ had been studied at QCD
NLO~\cite{Li:2008ym} and the partial NNLO result also been
given~\cite{Lansberg:2009db}. The hadroproduction of $\Upsilon+\gamma$ even proposed in reference~\cite{Dunnen:2014eta} as an ideal tool to investigate the transverse dynamics and polarization of gluon in proton. In our previous paper~\cite{Li:2014ava} we give a full result including the NLO
calculation on the color singlet and the color octet channel of the
$J/\psi$ associated production with a photon. We had shown that
several sets of LDMEs about $J/\psi$ production given different
prediction on $J/\psi+\gamma$. Some of them even result in the
unphysical distribution. How about the LDMEs on $\Upsilon$ which had
given well description of yield and polarization on $\Upsilon$
inclusive production at hadron colliders. In this paper we extend
our study to calculate the full results on $\Upsilon+\gamma$ at the
NLO.

According to the NRQCD factorization scheme, the differential cross
section for this process can be expressed as, \bea
&&\sigma(p+p \to \Upsilon+\gamma+X) =\sum_{i,j} \int dx_1dx_2 \\
&\times& G_p^i(x_1) G_{\bar{p}}^j(x_2) \hat{\sigma}(ij\to
(Q\bar{Q})_n+\gamma+X)\langle O^{\Upsilon}_n\rangle. \nonumber
\label{eqn:factorization} \eea The above equation indicate that the
cross section of this process is the convolution of the parton
distribution function $G_{p}^{i(j)}$ and the parton level short distance
coefficients $\hat{\sigma}$. The $\langle O^{J/\psi}_n\rangle$ are the
LDMEs of corresponding sub-processes. The parton level sub-process
are, \bea
g+g \to Q\bar{Q}\lbrack ^3S_1^1,^1S_0^8,^3S_1^8,^3P_J^8\rbrack +\gamma, \label{eqn:v1com}  \\
q+\bar{q} \to Q\bar{Q}\lbrack ^1S_0^8,^3S_1^8,^3P_J^8\rbrack +\gamma,\label{eqn:v2com}\\
g(q,\bar{q})+g(q,\bar{q}) \to Q\bar{Q}\lbrack
^3S_1^1,^1S_0^8,^3S_1^8,^3P_J^8\rbrack + \gamma +g. \label{eqn:rcom}
\label{eqn:r4com} \eea For the description of $J/\psi$ polarization
we adopt the usual definition in helicity frame as \bea
\alpha(p_t)=\frac{{\md\s_{11}}/{\md p_t}- {\md\s_{00}}/{\md
p_t}}{{\md\s_{11}}/{\md p_t}+{\md\s_{00}}/{\md p_t}}. \eea

As in our previous paper on the hadroproduction of
$J/\psi+\gamma$~\cite{Li:2014ava} we also use the same isolated
condition in reference~\cite{Frixione:1998jh} to isolate a photon
from the quark jet in some of the sub-processes. The isolated
condition is \bea p_t^i\le p_t^{\gamma}\frac{1-\cos
R_{\gamma_i}}{1-\cos \delta_0}~~~~for~~~~R_{\gamma_i} < \delta_0.
\eea The definitions of the $p_t^i$, $p_t^\gamma$, $\cos
R_{\gamma_i}$ and the $\delta_0$ can be found in
Ref.~\cite{Frixione:1998jh}. Here we set $\delta_0=$0.7. The newly
infrared divergence appearing in the calculation of p-wave processes
at the NLO is handled with the same method in
Ref.~\cite{Wang:2012tz}. We use the Feynman Diagram Calculation
(FDC) package~\cite{Wang:2004du} to generate the analytic results
and output the Fortran code for numerical evaluation.

Let us talk about the choice of parameters used in the numerical
calculation. In the calculation, the bottom quark mass is set as
4.73GeV and will vary from 4.63 to 4.83 to estimate the related
uncertainties. The factorization and the renormalization scales are
set as $\mu_r=\mu_f=\mu_0=\sqrt{(2m_b)^2+p_t^2}$ and will vary from
$\mu_0/2$ to $2\mu$ to estimate the uncertainties. The CTEQ6L amd
CTEQ6M PDFs are used in the calculation of the LO and NLO
convolutions and the $\alpha_s$ running in these PDFs are used to
calculate the cross section of the sub-processes. The center of mass
energy and cut condition for the $\Upsilon$ or the final photon are
set as $\sqrt{s}=7,8,14$TeV, $|y_{\Upsilon,\gamma}|\le3$,
$|\eta_{\gamma}| \le 1.45$ and $p_t^{\gamma} > 1.5,3,5,15$GeV.
The fine structure constant for the electromagnetic
coupling is chosen as $\frac{1}{128}$.
The different choice of the parameters are marked 
in the figures.

The extraction of the LDMEs in NRQCD is the key point to give the
rational theoretical predictions. As for the LDMEs for the
production of $\Upsilon$ there are many sets of them extracted by
two groups form Peking University (PKU)~\cite{Wang:2012is,Han:2014kxa} and the Institute of High Energy
Physics~\cite{Gong:2013qka,Feng:2015wka}. In
reference~\cite{Wang:2012is} the authors had given the three LDMEs
at the NLO for the first time. But they included the feed-down
contribution from $^3S_1^{[8]}$ channel of P-wave bottomonium in the
corresponding CO LDMEs of $\Upsilon$. Therefore, we do not use the
LDMEs in reference~\cite{Wang:2012is} in our numerical calculation.
The authors in reference~\cite{Gong:2013qka} extracted the LDMEs
related to the production of $\Upsilon$ individually by matching
their theoretical prediction at the NLO with the yield and the
polarization at the Tevatron and LHC. Thereafter, they updated their
analysis with three different schemes by including the newly
measured parameters, the mass of $\Upsilon(3S)$ and the fraction for
$\chi_{bJ}(3P) \to \Upsilon(3S)$~\cite{Feng:2015wka}. The PKU group
also extracted the individual LDMEs in the first version of
reference~\cite{Han:2014kxa} by matching the yield and the
polarization data. In the second version of
reference~\cite{Han:2014kxa} they decomposed the contribution of
P-wave color-octet subprocesses into the linear combination of the
two S-wave subprocesses. Therefore, just as in the $J/\psi$
case~\cite{Ma:2010vd} they extracted two linear combination of the
three LDMEs as \bea M^{\Upsilon}_{0,r_0}=\langle
O^{\Upsilon}(^1S_0^8)\rangle+\frac{r_0}{m^2_b}\langle
O^{\Upsilon}(^3P_0^8)\rangle,\label{eqn:M0}\\
M^{\Upsilon}_{1,r_1}=\langle
O^{\Upsilon}(^3S_1^8)\rangle+\frac{r_1}{m^2_b}\langle
O^{\Upsilon}(^3P_0^8)\rangle, \label{eqn:M1} \eea where $r_0$=3.8,
$r_1$=-0.52, $M^{\Upsilon}_{0,r_0}=13.70\times10^{-2}$GeV$^3$ and
$M^{\Upsilon}_{1,r_1}=1.17\times10^{-2}$GeV$^3$. Here we use the
same method as in our previous paper on $J/\psi+\gamma$. By
requiring the LDMEs to be positive we obtain two sets of LDMEs
referring to as "Han Extension" (Han1 and Han2) in the following parts. The seven
sets of LDMEs and related parameters, the $\mu_{\Lambda}$ and the
cuts on the transverse momentum of $\Upsilon$($P_t^{\Upsilon}$), are
listed in Table~\ref{tab:LDME}.

\begin{table}
\caption[]{The NRQCD LDMEs $\langle O^{\Upsilon}(n)\rangle $
extracted by two groups at the NLO with $\langle
O^{\Upsilon}(^3S_1^1)\rangle =$9.28 GeV$^3$.The other three
color-octet LDMEs are listed as following.(LDMEs in unit of
10$^{-2}$ GeV$^3$,$P_t^{\Upsilon}$ and $\mu_{\Lambda}$ in unit of
GeV.)}
\label{tab:LDME}
\renewcommand{\arraystretch}{1.5}
\[
\begin{array}{|c|c|c|c|c|c|}
\hline \hline  &O(^1S_0^8)& O(^3S_1^8)& O(^3P_0^8)/m_b^2&P_t^{\Upsilon}>&\mu_{\Lambda} \\
\hline \textrm{Han~\cite{Han:2014kxa}}&0.017&2.97&3.83&15&m_b\\
\hline \textrm{Han1~\cite{Han:2014kxa}}&0&3.04&3.61&15&m_b\\
\hline \textrm{Han2~\cite{Han:2014kxa}}&13.7&1.17&0&15&m_b\\
\hline \textrm{Gong~\cite{Gong:2013qka}}&11.15&-0.41&-0.67&8&m_bv\\
\hline \textrm{Feng1~\cite{Feng:2015wka}}&13.6&0.61&-0.93&8&m_bv\\
\hline \textrm{Feng2~\cite{Feng:2015wka}}&10.1&0.73&-0.23&8&m_bv\\
\hline \textrm{Feng3~\cite{Feng:2015wka}}&11.6&0.47&-0.49&8&m_b\\
\hline \hline
\end{array}
\]
\renewcommand{\arraystretch}{1.0}
\end{table}

\begin{figure}
\includegraphics[scale=0.35]{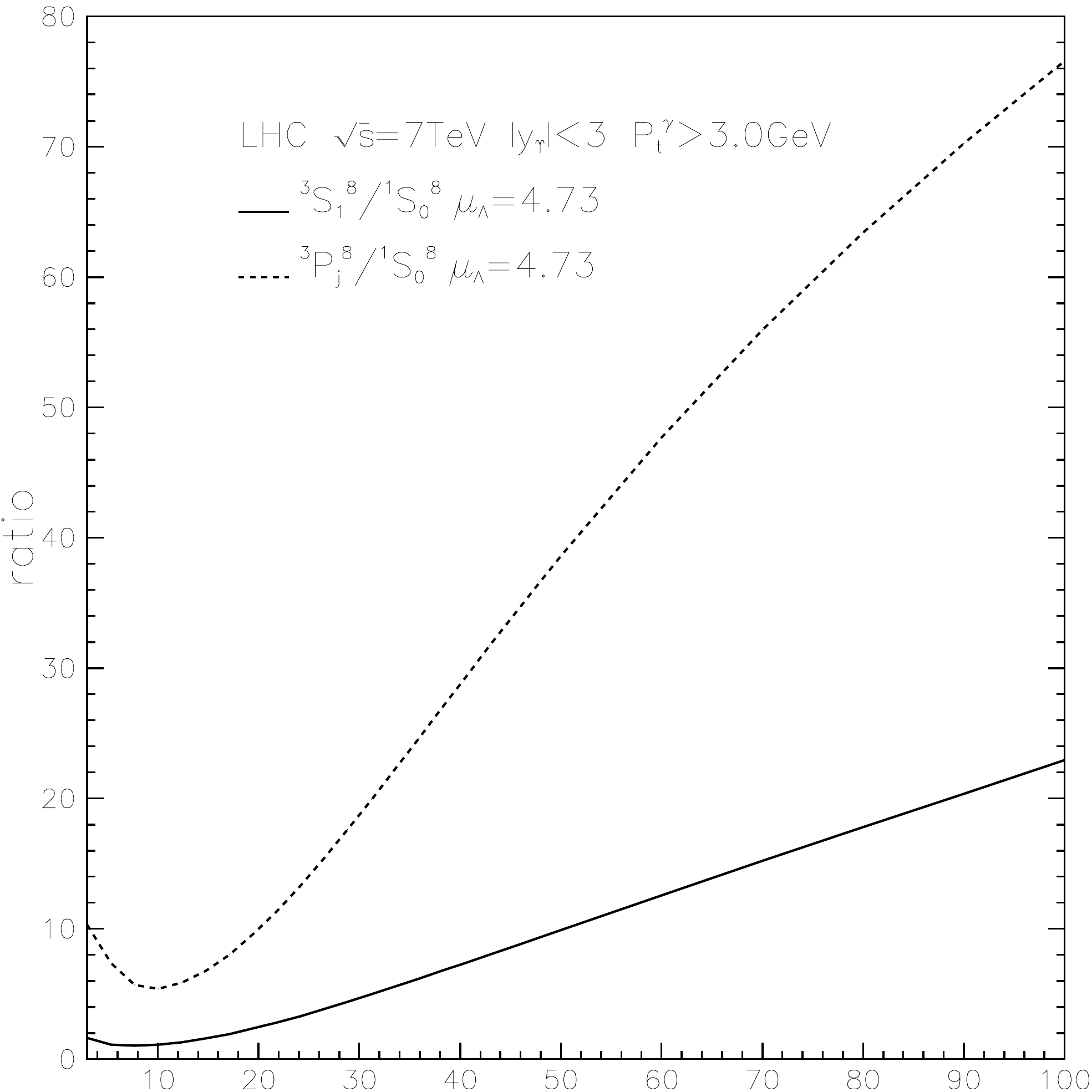}\\
\includegraphics[scale=0.35]{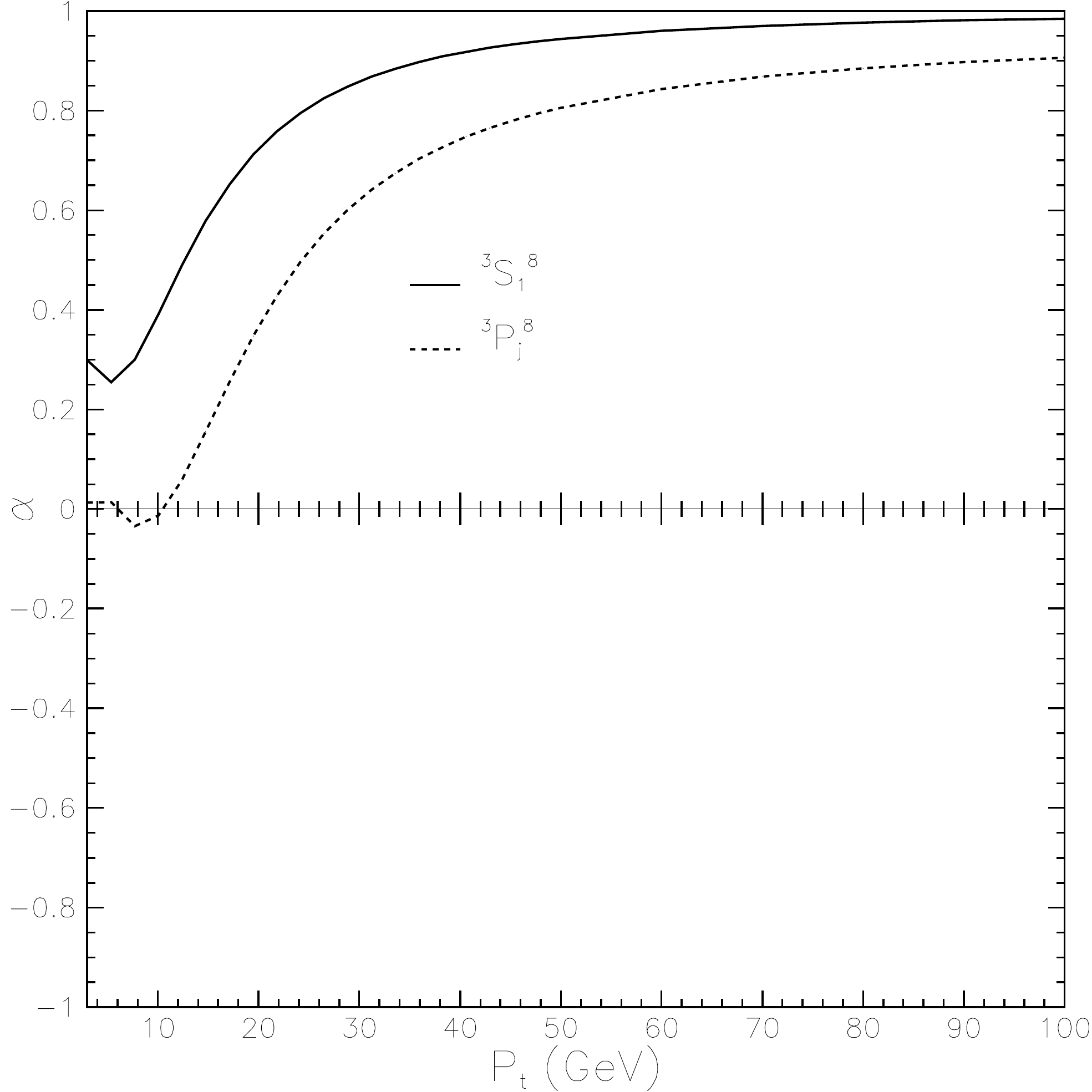}
\caption{\label{fig:pr7} The ratio of
$d\sigma(^3P_J^8)/d\sigma(^1S_0^8)$ and
$d\sigma(^3S_1^8)/d\sigma(^1S_0^8)$ as functions of $P_t$.}
\end{figure}

In figure~\ref{fig:pr7} we show some features of the sub-processes.
The ratios between the three sub-processes are plotted in the upper
parts of the figure~\ref{fig:pr7}. We can see that with the
increase of transverse momentum the cross sections for the $^3S_1^8$
and $^3P_J^8$ sub-processes at the parton level become muche larger than
that of the $^1S_0^8$ sub-process and there is no linear correlation
among them, just like the associated production of $J/\psi+\gamma$. 
Because the $^1S_0^8$
sub-process is unpolarized we just show the polarization of the
other two sub-processes in the lower part of figure~\ref{fig:pr7}.
Both of the previous two sub-processes show the transverse
polarization in almost the whole $P_t$ region. This two figures
together with the choice of the LDMEs will give qualitative hints on
the polarization of the final $\Upsilon$.

Because the contribution from $^3S_1^8$ and $^3P_J^8$ subprocesses
is dominant the numerical results with the Han set is almost the
same as that of the Han1. We just show the results of two
extension cases. There is similar situation between the Gong and the
Feng1 sets. The sets Feng1 and Feng3 give the similar trend on the
yield and the polarization, which become unphysical in the large
$p_t$ region. Therefore, we just plot the numerical results with the
LDMEs sets Han extension1, Han extension2, Feng1 and Feng2 in the
following parts in our paper.

\begin{figure*}
\begin{center}
\includegraphics[scale=0.29]{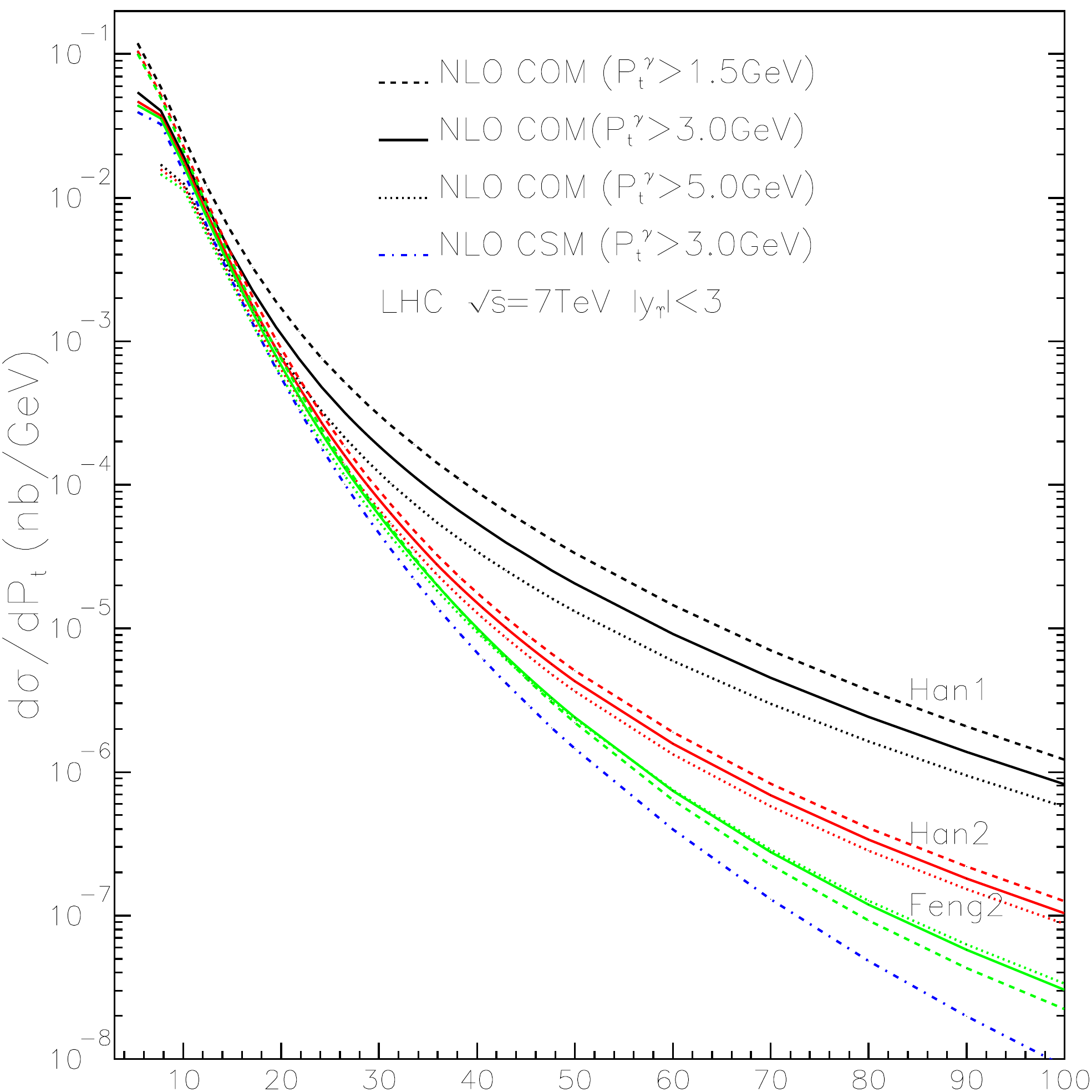}
\includegraphics[scale=0.29]{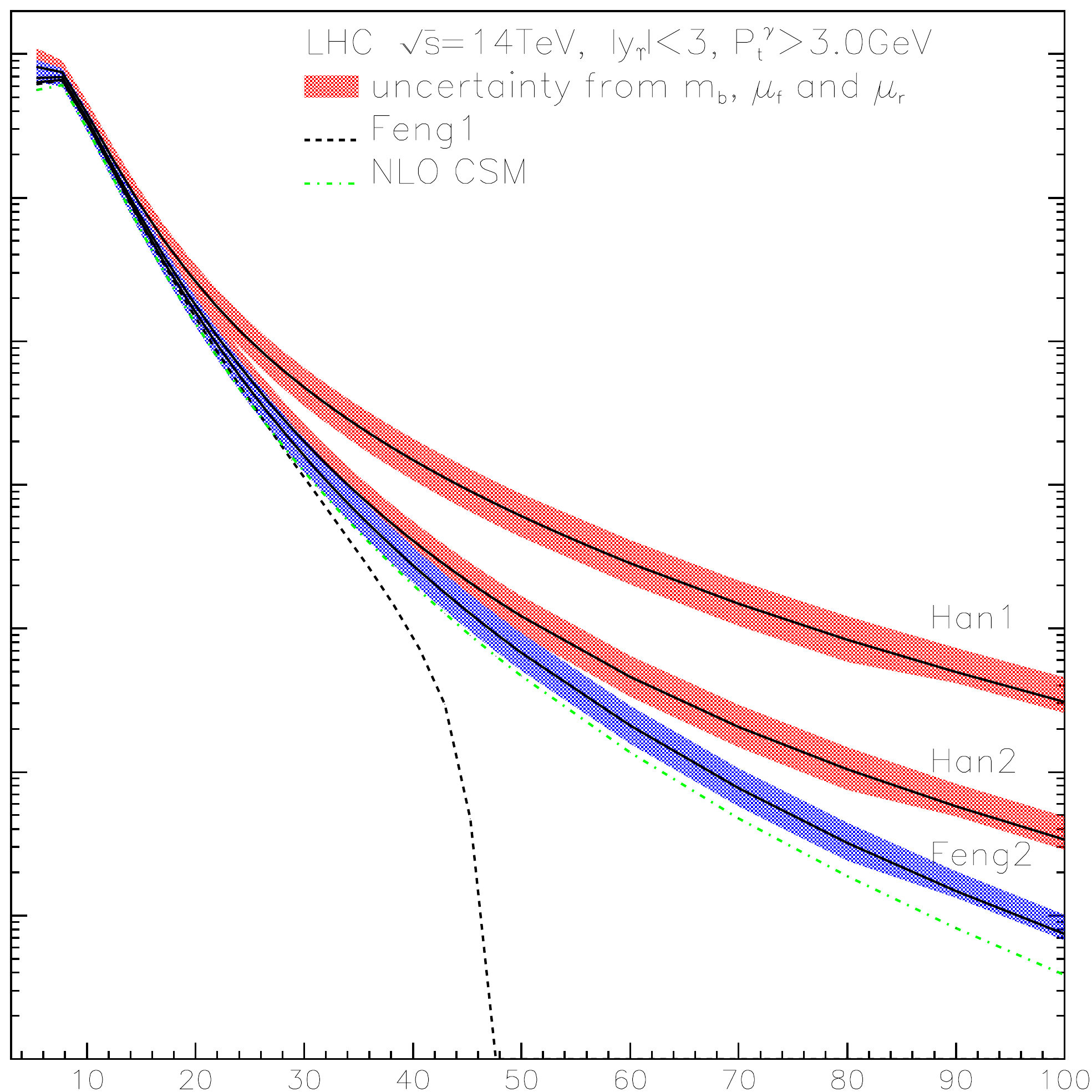}
\includegraphics[scale=0.29]{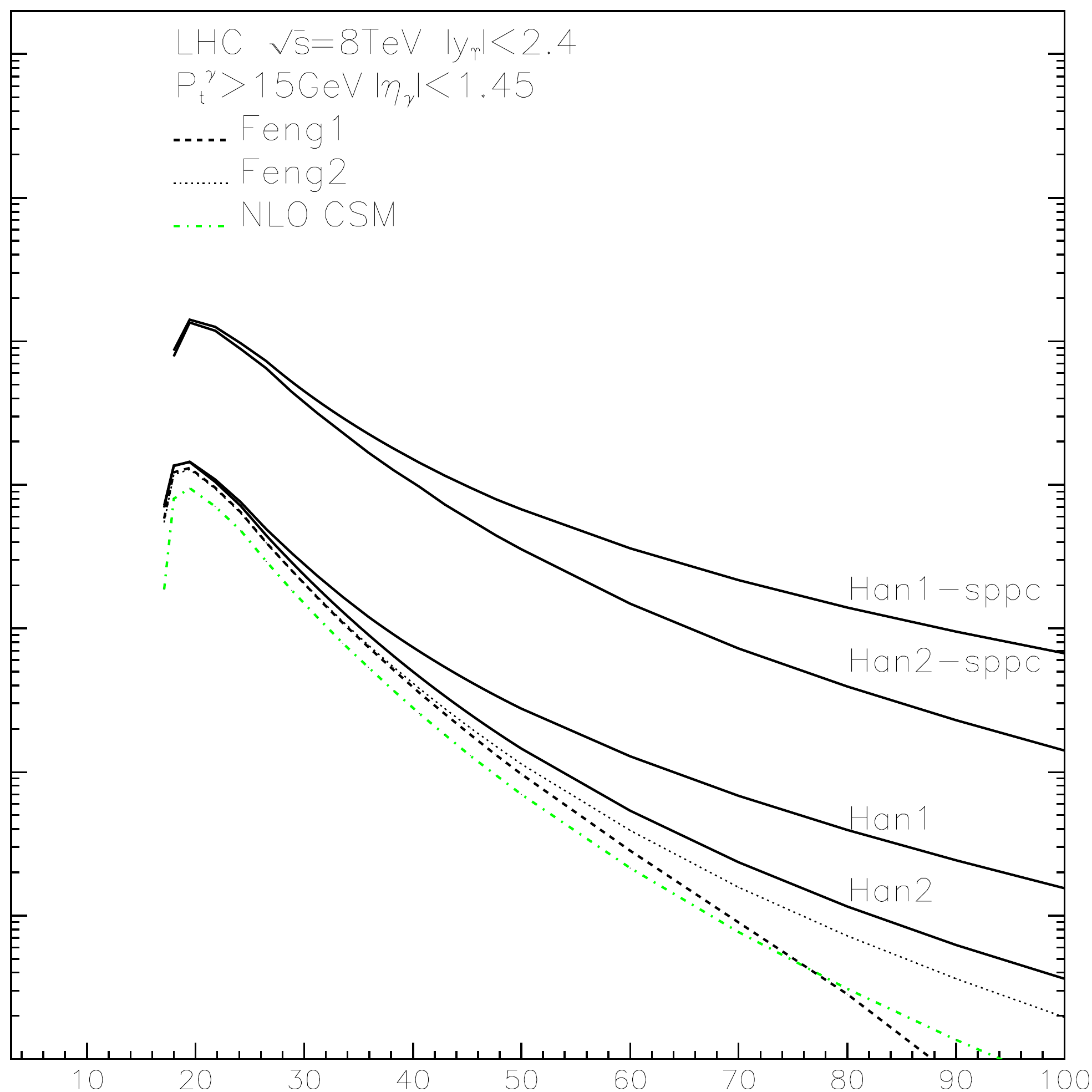}
\includegraphics[scale=0.29]{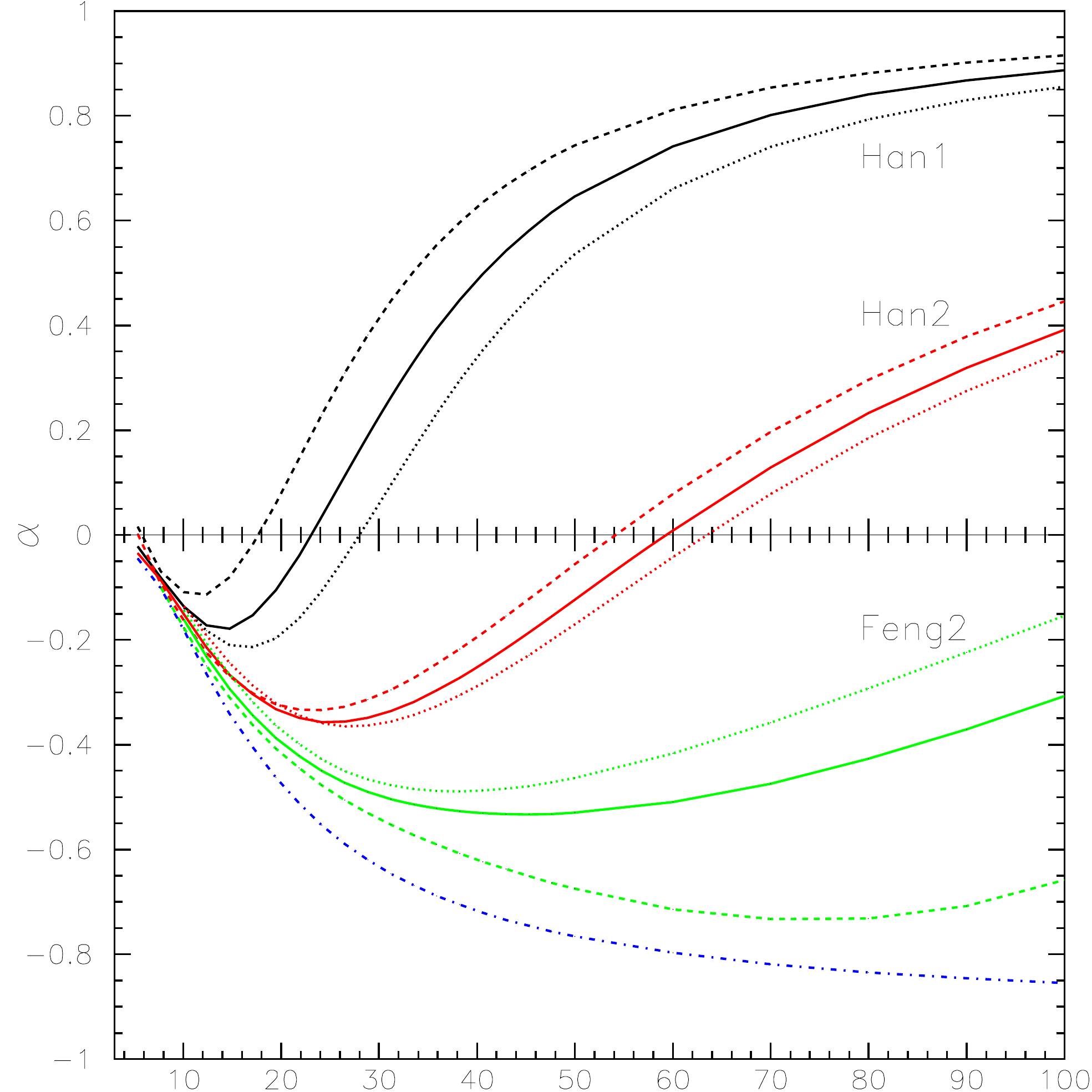}
\includegraphics[scale=0.29]{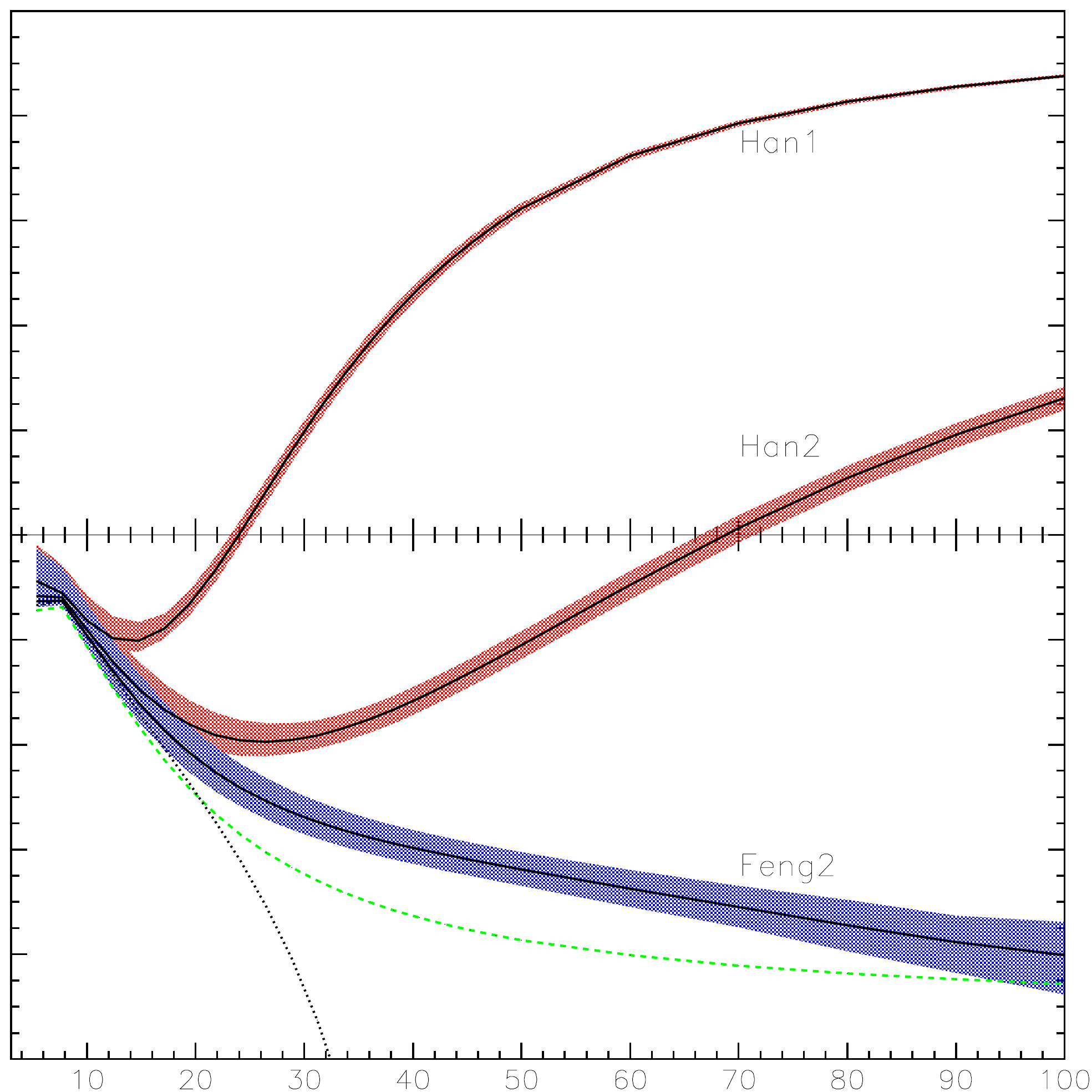}
\includegraphics[scale=0.29]{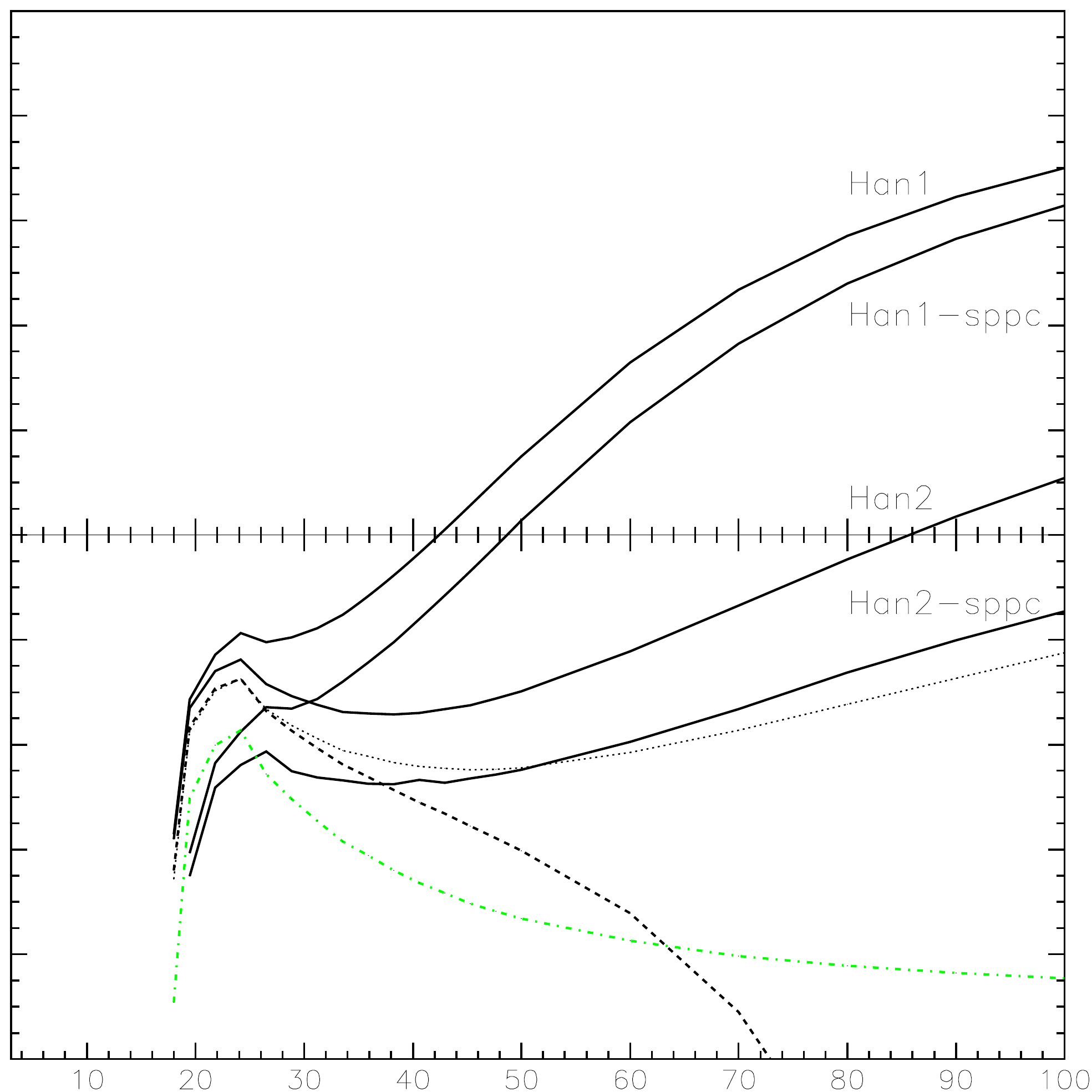}
\caption{\label{fig:ptdis} The $p_t(\Upsilon)$ distributions for
$\Upsilon+\gamma$ production (upper parts) and polarization (lower
parts) with different conditions. Figures in the same column are of
the same conditions and line types. The shaded band in the second
column represent the uncertainty from variation of $\mu_f$ \&
$\mu_r$ and $m_b$. The third column shows results with
$p_t^{\gamma}>15$GeV and different LDME sets.}
\end{center}
\end{figure*}

We plot three sets of results with different cut conditions
and observables in three columns in figure~\ref{fig:ptdis}. In the
first column we investigate the dependence of cross section and the
polarization on the $p_t$ cuts of the associated photon with three
sets of LDMEs which can give the physical results on the $p_t$
distribution and  the polarization. We choose three $p_t^{\gamma}$
cuts and plot the corresponding curves. Comparing to the CSM results
at the NLO even at the large $p_t$ region(about 100GeV) the cross
section and the polarization do not have a manifestly convergence
behavior in the $p_t$ region being studied. This is due to the
infrared divergence related to the photon which we handled by using
the isolated condition. They almost have a parallel shift with
different $p_t^{\gamma}$ cuts.

The second column in figure~\ref{fig:ptdis} shows our results at
$\sqrt{s}=14$GeV. The shaded band presents the dependence of physical
observables on three parameters, bottom quark mass $m_b$,
renormalziation scale $\mu_r$ and the factorization scale $\mu_f$.
The numerical results show the similar feature as that in the
$J/\psi$ case~\cite{Li:2014ava}. At first by including the
contribution of color-octet processes the cross section for the
inclusive hadroproduction of $\Upsilon+\gamma$ is enhanced about 2
or 70 times larger than that of the CSM result at NLO with the three
sets of LDMEs which can give physical results. The polarization of
$\Upsilon$ changes from mainly longitudinal in CSM to slightly longitudinal, slightly transverse and even mainly transverse at the NLO
with the three sets of LDMEs. The uncertainty of production rate
becomes larger with the increase of $p_t$ and at the same time the
polarization parameter exhibit the more sophisticated feature. The
polarization parameter obtained by using the Han extension sets of
LDMEs converges with the increasing of $p_t$ and the one with the Feng2
set of LDMEs does not. From the values of the LDMEs we can infer
that the uncertainty mainly comes from the $^1S_0^8$ channel. The
results with the Feng1 set of LDMEs are also plotted in the figures,
which give negative cross section in the $p_t$ distribution at about
$p_t=48$GeV and result a polarization parameter lower than -1 when
$p_t>32$GeV.

Just as in the case of $J/\psi+\gamma$ the associated production
of $\Upsilon+\gamma$ provides an opportunity to distinguish or to assess the
different sets of LDMEs obtained with different schemes on
$\Upsilon$ production. We calculate the theoretical prediction for
the yield and the polarization of $\Upsilon$ in this process with
the $\sqrt{s}=8$GeV. The large hadron collider (LHC) has 23$fb^{-1}$
integrated luminosity at $\sqrt{s}=8$GeV. We choose $p_t^{\gamma}
>15$GeV to suppress the background and $\eta_{\gamma}<1.45$ for
photon reconstruction efficiency consideration. The numerical
results show that in the large $p_t$ region the differential cross
section which including the COM contribution is about 2 to 20 times
larger than the CSM result. The ratio is smaller than that of the $J/\psi+\gamma$. The Feng1 set of LDMEs also gives the unphysical predictions
on the $p_t$ yield and the polarization. The photon reconstruction
efficiency under the above pseudo-rapidity and $p_t$ cut condition
is 0.7~\cite{CMS:2013aoa}. The branch ratio of $\Upsilon$ di-lepton
decay channel is $Br(\Upsilon \to \mu^+\mu^- \text{and} e^+ e^-)=0.05$.
Therefore we can expect there are 0.5$\sim$5 events at
$p_t^{\Upsilon}=100$GeV or 9$\sim$30 events at
$p_t^{\Upsilon}=60$GeV with $\sqrt{s}=8$TeV. If we raise the center
of mass energy to 100TeV and keep the other conditions unchanged the
event number can be enhanced to 32$\sim$161 or 322$\sim$782
respectively.

In summary, we investigate the inclusive production of
$\Upsilon+\gamma$ at hadron collider. By including the three
different color-octet sub-processes we give complete theoretical
predictions on the yield and the polarization at the QCD NLO within
NRQCD framework. Comparing to our previous work~\cite{Li:2014ava}
the cross section for this process is enhanced several times or
even two orders in the large $p_t$ region. The inclusion of
color-octet sub-processes at the NLO changes the polarization from
longitudinal one to transverse one. Although the scale chosen and
the variation of the bottom quark mass can bring the uncertainties
to some extent the main uncertainty of the theoretical prediction
comes from the difference among the LDMEs. We investigate seven sets
of LDMEs and only four of them, Han, Han extension1, Han
extension2,and feng2 can give the physical predictions on the yield
and the polarization in the $p_t$ region that we studied. From
figure~\ref{fig:pr7} it can be seen that the p-wave color-octet
sub-process give the dominate contribution in the large $p_t$
region. Therefore we infer that any set of LDMEs with negative
$O(^3P_0^8)$ will results in the negative prediction on the yield at
sufficiently large $p_t$ region and the Feng2 set may meet this
problem with the increase of $p_t$. The estimate with the LHC integrated luminosity
at $\sqrt{s}=8$TeV indicate that it is not optimistic to investigate
this process at the LHC. The result with $\sqrt{s}=100$TeV shows us
this process can be studied in the next generation hadron colliders,
such as the Super Proton-Proton Collider(SPPC).

We acknowledge the supports from the National Natural Science
Foundation of China under Grants No. 11105152, 11375137 and U1832160, the Natural Science Foundation of Shaanxi Province under Grants No.
2015JQ1003 and the Fundamental Research Funds for the Central
Universities.


\end{document}